\documentclass[pra,twocolumn,amsmath,amssymb,floatfix]{revtex4}
\usepackage{graphicx}
\usepackage{dcolumn} 
\usepackage{bm}
\usepackage{amsmath}
\usepackage{latexsym}
\usepackage{amsfonts}
\usepackage{dcolumn}
\usepackage{color}
\usepackage{amssymb}
\usepackage[rightcaption]{sidecap}
\newcommand{\be}{\begin{equation}}
\newcommand{\ee}{\end{equation}}
\newcommand{\bea}{\begin{eqnarray}}
\newcommand{\eea}{\end{eqnarray}}
\newcommand{\nn} {\nonumber}
\renewcommand{\vr} {{\bf r}}

\def\a{\alpha}

\def\g{\gamma}

\def\d{\delta}
\def\D{\Delta}

\def\ve{\varepsilon}

\def\r{\rho}

\def\S{\Sigma}

\def\vf{\varphi}

\def\w{\omega}

\def\bra{\langle}
\def\ket{\rangle}

\def\xc{{\rm xc}}
\def\x{{\rm x}}

\begin{document}
\title{Discontinuities of the exchange-correlation kernel and charge-transfer excitations in time-dependent density functional theory}
\author{Maria Hellgren}
\author{E. K. U. Gross}
\affiliation{Max-Planck-Institut f\"ur Mikrostrukturphysik, Weinberg 2, D-06120 Halle, Germany}  
\date{\today}
\begin{abstract}
We identify the key property that the exchange-correlation (XC) kernel of 
time-dependent density functional theory must have in order to describe 
long-range charge-transfer excitations. We show that the discontinuity of the 
XC potential as a function of particle number induces a space -and frequency-dependent discontinuity of the XC kernel which diverges as $r\to\infty$. In a combined donor-acceptor system, the same discontinuity compensates for the vanishing overlap between the acceptor and donor orbitals, thereby yielding a finite correction to the Kohn-Sham eigenvalue differences. This mechanism is illustrated to first order in the Coulomb interaction.
\end{abstract}
\maketitle
\section{Introduction}
The theoretical prediction of the excitation spectra of interacting electronic systems is a major challenge in quantum chemistry and condensed matter physics. A method that has been gaining popularity in the past years is time-dependent density functional theory (TDDFT) offering a rigorous and computationally efficient approach for treating excited states of large molecules and nanoscale systems. 
In TDDFT the interacting electronic density is calculated from a system of non-interacting electrons moving in an effective local Kohn-Sham (KS) potential \cite{KS}. 
The KS potential is the sum of the external, the Hartree and the so-called exchange-correlation (XC) potential, $v_\xc(\vr t)$, which, due to the Runge-Gross theorem \cite{rg0}, is a functional of the density. In the linear response regime, the excitation energies can be extracted from the poles of the linear density response function. As a consequence, given the variational derivative $f_\xc(\vr t,\vr' t')=\d v_{\xc}(\vr t)/\d n(\vr' t')$, also known as the XC kernel \cite{grosskohn}, it is possible to formulate an RPA-like equation for the exact excitation spectrum \cite{pgg96,casbook}. In practical calculations, this equation is solved using some approximate XC potential and kernel where the most popular ones are based on the adiabatic local density approximation (ALDA), leading to kernels local in both space and time. Despite this simple structure optical excitations of small molecules are successfully predicted. However, several shortcomings have also been reported: excitons in solids are not captured, \cite{reining,marini} double excitations are missing \cite{m1} and charge-transfer (CT) excitations are qualitatively incorrect \cite{dreuw,mt06,handy,roi1}. In this work we will be concerned with the last problem and to see why ALDA fails in this case, we consider a charge transfer between two neutral Coulombic fragments. The asymptotic limit of the excitation energy is then given by
\be
\w_{\rm CT}=I_d-A_a-1/R,
\ee 
where $I_d$ is the ionization energy of the donor, $A_a$ is the affinity of the acceptor and $R$ is their separation. 
In TDDFT the starting point is the KS system which yields the exact $I_d$ but only an approximate $A_a$. Thus, the XC kernel must both account for the $1/R$ correction and shift the KS affinity. 
The linear response equations involve, however, only matrix elements of $f_\xc$ between so-called excitation functions $\Phi_{ia}(\vr)=\vf_i(\vr)\vf_a(\vr)$, i.e., products of occupied and unoccupied KS orbitals. As the distance between the fragments increases these products vanish exponentially and thus there is no correction to the KS eigenvalue differences unless the kernel diverges \cite{barct,tozer2}. Kernels from the ALDA, or adiabatic GGA's for that matter, do not contain such divergency and it is as yet not understood how this extreme behavior should be incorporated in approximate functionals. 

Whenever two subsystems are spatially well-separated it is possible to treat one of the subsystems in terms of an ensemble containing states with different number of particles. DFT has been generalized to non-integer charges and as an important consequence it was found that the XC potential jumps discontinuously by a constant at integer particle numbers in order to align the highest occupied KS eigenvalue with the chemical potential \cite{pplb82}. In this way, the true affinity, $A$, is equal to the sum of the KS affinity, $A_s$, and the discontinuity. Not surprisingly, it has therefore been argued that the discontinuity must play an important role in describing CT excitations within TDDFT \cite{vcu09}. 

So far, only the XC potential has been the target for investigating discontinuities in DFT and TDDFT \cite{lein,burketran,gl}. In this work we instead examine possible discontinuities of the XC kernel. We demonstrate the existence of a discontinuity and we study its properties. Furthermore, we give an explicit example through a numerical study of the EXX functional. Finally, as a first application, we demonstrate the crucial role of the discontinuity for capturing CT excitations in linear response TDDFT.
\section{Derivative discontinuity in DFT} 
We start by considering a static system of electrons described by a statistical operator $\hat{\r}=\sum_k\alpha_k|\Psi_k \ket\bra \Psi_k|$, where $|\Psi_k\ket$ denotes the ground state of $k$ particles corresponding to the Hamiltonian $\hat{H}=\hat{T}+\hat{V}+\int d\vr\, w(\vr)\hat{n}(\vr)$, in which $\hat{T}$ is the kinetic energy, $\hat{V}$ the inter-particle interaction and $w$ is the external potential. 
The ground-state energy $E_0$ of the system with average number of particles $N$ is obtained by minimizing the functional $E_w[n]=F[n]+\int d\vr \, w(\vr) n(\vr)$, where 
$F[n]=\min_{\hat{\r}\to n}{\rm Tr}[\hat{\r}(\hat{T}+\hat{V})]$, under the constraint that $N=\int d\vr \,n(\vr)$. At the minimum $n=n[w,N]$ coincides with the ground-state density. The XC energy is defined as $E_\xc[n]=F[n]-T_s[n]-U[n]$, where $T_s[n]=\min_{\hat{\r}\to n}{\rm Tr}[\hat{\r}\hat{T}]$ is the non-interacting kinetic energy and $U[n]$ is the Hartree energy. Assuming that the density can be reproduced by an ensemble of non-interacting electrons the XC part of the KS potential is given by $v_{\xc}(\vr)=\d E_{\xc}/\d n(\vr)$. In general $E_0$ and, in particular, $E_\xc[n[w,N]]$ has derivative discontinuities at integer particle numbers $N_0$ \cite{perlev,ss83}. 
The partial derivative of $E_\xc$ with respect to $N$,
\bea
\frac{\partial E_{\xc}}{\partial N}=\int d \vr\,\,  v_{\xc}(\vr)\frac{\partial n(\vr)}{\partial N},
\label{forstaderivN}
\eea 
has two sources of discontinuous behavior: (i) the quantity
\be
f(\vr)\equiv\frac{\partial n(\vr)}{\partial N},
\ee 
known as the Fukui function, may have different right and left limits, $f^+$ and $f^-$ (the superscript $\pm$ refers to the value of the quantity at $N=N_0+0^\pm$), (ii) the XC potential may be discontinuous, $v^+_{\xc}(\vr)=v^-_{\xc}(\vr)+\D_\xc$, where $\D_\xc$ is a constant. Below we will show that also the second variation of $E_\xc$ with respect to the density, the static XC kernel $f_\xc(\vr,\vr')=\d v_\xc(\vr)/\d n(\vr')$, has discontinuities which are related to derivative discontinuities in the density itself. As pointed out in previous work \cite{hvb2,awst3,casbook}, the {\em particle number conserving} density response is unaffected by adding to $f_\xc$ a function depending only on one of the coordinates. In line with the results of Ref. \onlinecite{gal1} we therefore argue that the discontinuities of $f_\xc$ must be of the form 
\be
f^+_\xc(\vr,\vr')=f^-_\xc(\vr,\vr')+g_\xc(\vr)+g_\xc(\vr').
\ee 
In the following we will show a simple procedure for determining $\D_\xc$ and $g_\xc(\vr)$ which is useful whenever $E_\xc$ is an implicit functional of the density via, e. g., the KS Green function. 
\subsection{XC potential}
For $N> N_0$, we write $v_\xc(\vr)=v^-_\xc(\vr)+\D_\xc(\vr)$ and cast Eq. (\ref{forstaderivN}) into 
\bea
\int d\vr\,\, \D_\xc(\vr)f(\vr)=\frac{\partial E_{\xc}}{\partial N}-\int d \vr\, v^-_{\xc}(\vr)f(\vr).
\label{muxc}
\eea
In the limit $N\rightarrow N_{0}^+$, $\D_\xc(\vr)\to\D_\xc$ and we find a formal expression for the discontinuity of $v_\xc$
\bea
\D_\xc=\left.\frac{\partial E_{\xc}}{\partial N}\right|_+-\int d \vr\, v^-_{\xc}(\vr)f^+(\vr).
\label{muxc2}
\eea
This expression can be used as the starting point for deriving the well-known MBPT-formula for the correction to the gap \cite{perde}, as we will now demonstrate. From the Klein functional within MBPT \cite{klein} it is possible to construct an XC energy functional in terms of the KS Green function $G_s(\vr,\vr',\w)$ \cite{vbdvls05}. In this case $\S_\xc=\d E_\xc/\d G_s$, where $\S_\xc$ is the self-energy evaluated at $G_s$. The derivative of $E_\xc$ with respect to $N$ is then given by
\bea
\frac{\partial E_\xc}{\partial N}=\int \frac{d\w}{2\pi}\, \int d\vr d\vr' \S_{\xc}(\vr,\vr',\w)\frac{\partial G_s(\vr,\vr',\w)}{\partial N}.
\label{dedg}
\eea
In order to evaluate the derivative of $G_s$ with respect to $N$ we consider an ensemble described by a spin-compensated mixture of states with electron number $N_0$ and $N_0+1$. The KS ensemble Green function for $N\in [N_0,N_0+1]$ is given by  
\bea
G_s(\vr,\vr',\w)&=&\sum_{k=1}^{N_0}\frac{\vf_k(\vr)\vf_k(\vr')}{\w-\ve_k-i\eta}+\sum_{k=N_0+2}^{\infty}\frac{\vf_k(\vr)\vf_k(\vr')}{\w-\ve_k+i\eta}\nn\\
&&\!\!\!\!\!\!\!\!\!\!\!+\frac{p}{2}\frac{\vf_{\rm L}(\vr)\vf_{\rm L}(\vr')}{\w-\ve_{\rm L}-i\eta}+\left(1-\frac{p}{2}\right)\frac{\vf_{\rm L}(\vr)\vf_{\rm L}(\vr')}{\w-\ve_{\rm L}+i\eta}
\eea
where $p=N-N_0$ and the subscript ${\rm L}$ signifies the lowest unoccupied molecular orbital (LUMO) of the KS system, which is considered partially occupied and partially unoccupied. Notice that the KS orbitals $\vf_k$ and eigenvalues $\ve_k$ also depend on $N$ via the KS potential $V_s$. The derivative of $G_s$ with respect to $N$ is now easily carried out
\bea
\frac{\partial G_s(\vr,\vr',\w)}{\partial N}&=&\frac{1}{2}\frac{\vf_{\rm L}(\vr)\vf_{\rm L}(\vr')}{\w-\ve_{\rm L}-i\eta}-\frac{1}{2}\frac{\vf_{\rm L}(\vr)\vf_{\rm L}(\vr')}{\w-\ve_{\rm L}+i\eta}\nonumber\\
&&+\int d\vr_1\frac{\d G_s(\vr,\vr',\w)}{\d V_s(\vr_1)}\frac{\partial V_s(\vr_1)}{\partial N}.
\label{muxc3}
\eea
From Eq. (\ref{muxc3}) and Eq. (\ref{dedg}) we find
\bea
\left.\frac{\partial E_\xc}{\partial N}\right|_+&=&\int d\vr d\vr' \vf^+_{\rm L}(\vr)\S^+_{\xc}(\vr,\vr',\ve_{\rm L})\vf^+_{\rm L}(\vr')\nn\\
&&\!\!\!\!\!\!\!\!\!\!\!\!\!\!\!\!\!\!\!\!\!\!\!\!\!\!\!+\int d\vr d\vr'd\vr_1\S^+_{\xc}(\vr,\vr',\w)\left.\frac{\d G_s(\vr,\vr',\w)}{\d V_s(\vr_1)}\frac{\partial V_s(\vr_1)}{\partial N}\right|_+
\label{sig}
\eea
The second term on the right hand side of Eq. (\ref{muxc2}) can be written as
\bea
\int d \vr\, v^-_{\xc}(\vr)f^+(\vr)&=&\int d \vr\, v^-_{\xc}(\vr)|\vf^+_{\rm L}(\vr)|^2\nn\\
&&\!\!\!\!\!\!\!\!\!\!\!\!\!\!\!\!\!\!\!\!\!\!\!\!\!\!\!\!\!\!\!+\int d\vr'd\vr_1\left.v^+_{\xc}(\vr')\frac{\d n(\vr')}{\d V_s(\vr_1)}\frac{\partial V_s(\vr_1)}{\partial N}\right|_+.
\label{vxi}
\eea
The discontinuity $\D_\xc$ is now easily determined. The second terms on the right hand side of Eqs. (\ref{sig}-\ref{vxi}) will cancel by virtue of the linearized Sham-Schl\"uter equation \cite{ss} and we find
\bea
\D_\xc=\int d\vr d\vr' \vf_{\rm L}(\vr)\S^+_{\xc}(\vr,\vr',\ve_{\rm L})\vf_{\rm L}(\vr')\nn\\
-\int d \vr\, v^-_{\xc}(\vr)|\vf_{\rm L}(\vr)|^2,
\label{muxc4}
\eea
where we have omitted the superscript on the orbitals since they are continuous with respect to $N$. Eq. (\ref{muxc4}) agrees with the one in Ref. \onlinecite{perde}.
\subsection{XC kernel}
Next, we turn to the XC kernel for which we exhibit the discontinuities by taking the functional derivative of $\partial E_\xc/\partial N$ (Eq. (\ref{forstaderivN})) with respect to $w$. This yields
\bea
\frac{\d}{\d w(\vr_1)}\frac{\partial E_\xc}{\partial N}&=&\!\int\! d \vr d \vr'\,\chi(\vr_1,\vr') f_{\xc}(\vr',\vr)f(\vr)\nn\\
&&+\!\int \!d \vr \,
v_\xc(\vr)\frac{\d f(\vr)}{\d w(\vr_1)}
\label{muxcdeltan1}
\eea
where we have used the chain rule 
\be
\frac{\d v_\xc(\vr)}{\d w(\vr_1)}=\int d\vr'\frac{\d v_\xc(\vr)}{\d n(\vr')}\frac{\d n(\vr')}{\d w(\vr_1)}
\ee 
and identified the linear density response function $\chi(\vr_1,\vr')=\d n(\vr')/\d w(\vr_1)$. Then, we write $f_\xc(\vr,\vr')=f^-_\xc(\vr,\vr')+g_\xc(\vr,\vr')$, insert in Eq. (\ref{muxcdeltan1}), and take the limit $N\rightarrow N_{0}^+$. According to the discussion above, in this limit $g_\xc(\vr,\vr')\to g_\xc(\vr)+g_\xc(\vr')$. Using furthermore that $\int d\vr\,\chi(\vr,\vr')=0$ and $\int d\vr f(\vr)=1$ we find the following equation for the discontinuity $g_\xc$ of $f_\xc$:
\begin{widetext}
\bea
 &&\!\!\!\!\!\!\!\!\!\!\!\int d \vr \,\chi(\vr_1,\vr) g_\xc(\vr)=\left.\frac{\d}{\d w(\vr_1)}\frac{\partial E_\xc}{\partial N}\right|_+-\!\int\! d \vr d \vr'\,\chi(\vr_1,\vr) f^-_{\xc}(\vr,\vr')f^+(\vr')-\!\int \!d \vr \,
v^+_\xc(\vr)\left.\frac{\d f(\vr)}{\d w(\vr_1)}\right |_+
\label{muxcdeltan}
\eea
We notice that Eq. (\ref{muxcdeltan}) only determines $g_\xc$ up to a constant. This constant can, however, easily be fixed by considering $\partial^2 E_\xc/\partial N^2$ in the limit $N\rightarrow N_{0}^+$
\bea
 2\int d \vr \,f^+(\vr)g_\xc(\vr)=\left.\frac{\partial^2 E_\xc}{\partial N^2}\right|_+-\int  d \vr' \,v^+_\xc(\vr')\left.\frac{\partial f(\vr')}{\partial N}\right|_+ -\int d \vr d\vr' \, f^+(\vr)f^-_{\xc}(\vr,\vr')f^+(\vr').
\label{muxcdeltandeltan}
\eea
\end{widetext}
This equation does not allow an arbitrary constant in $g_\xc$. Consequently, Eq. (\ref{muxcdeltan}) and Eq. (\ref{muxcdeltandeltan}) together uniquely determines the discontinuity of $f_\xc$.

To gain insight on the $\vr$-dependence of $g_\xc(\vr)$ we employ a common energy denominator approximation (CEDA) \cite{kli} to Eq. (\ref{muxcdeltan}). To do so, we first notice that the derivative with respect to $w$ can be replaced by the derivative with respect to $V_s$ using the chain-rule since the density is a functional of $w$ only via $V_s$. The CEDA then allows us to partially invert the KS response function $\chi_s$ analytically. We will focus on the left hand side of Eq. (\ref{muxcdeltan}) and on the last term on the right hand side. These terms are less sensitive to the approximation used for $E_\xc$ and should therefore give rise to a general behavior. If all energy denominators are set to the constant $\D\epsilon$ we find on the left hand side of Eq. (\ref{muxcdeltan}) 
\bea
\int d \vr \,\chi_s(\vr_1,\vr) g_\xc(\vr)&\approx&-\frac{2}{\D\epsilon} n(\vr_1)g_\xc(\vr_1)\nn\\
&&\!\!\!\!\!\!\!\!\!\!\!\!\!\!\!\!\!\!\!\!\!\!+\frac{2}{\D\epsilon} \!\int d\vr\,\g(\vr_1,\vr)g_\xc(\vr)\g(\vr,\vr_1)
\label{cedag}
\eea 
where $\g$ is the KS density matrix. If we focus on the last term in Eq. (\ref{muxcdeltan}) and use $f^+(\vr)\approx |\vf_{\rm L}(\vr)|^2$ we find in the CEDA
\bea
\!\int \!d \vr \,
v^+_\xc(\vr)\left.\frac{\d f(\vr)}{\d V_s(\vr_1)}\right |_+\approx-\frac{2}{\D\epsilon} |\vf_{\rm L}(\vr_1)|^2 v_\xc^-(\vr_1)\nn\\
\!\!\!\!\!\!\!+\frac{2}{\D\epsilon} |\vf_{\rm L}(\vr_1)|^2 \int d\vr |\vf_{\rm L}(\vr)|^2 v_\xc^-(\vr)\nn\\
+\frac{4}{\D\epsilon}\vf_{\rm L}(\vr_1)\int d\vr\,\g(\vr_1,\vr)\vf_{\rm L}(\vr)v_\xc^-(\vr)
\label{cedavx}
\eea
From Eq. (\ref{cedag}) and Eq. (\ref{cedavx}) we can extract an approximate asymptotic behavior
\be
g_\xc(\vr)\sim -\frac{|\vf_{\rm L}(\vr)|^2}{n(\vr)}v_\xc(\vr)\sim e^{2(\sqrt{2I}-\sqrt{2A_s})\, r},
\label{gxcdiv}
\ee
where $I$ is the ionization energy \cite{ab} and $A_s$ the KS affinity. Thus, we can conclude that if $I>A_s$ $g_\xc$ contains a term which diverges exponentially as $r\to\infty$. That the first term of Eq. (\ref{muxcdeltan}) would exactly cancel the term of Eq. (\ref{gxcdiv}) is highly unlikely and we will explicitly see that within an approximation that accounts for the derivative discontinuity as the EXX approximation this is not the case. To obtain Eq. (\ref{gxcdiv}) we have used the CEDA but we will show below that the discontinuity obtained from the full solution of Eq. (\ref{muxcdeltan}) exhibits the same behavior. In addition, we will demonstrate that this feature is responsible for sharp peak structures as well as divergences in the kernel of a combined donor-acceptor system. 
\section{Discontinuity of the dynamical XC kernel} 
So far the analysis has been limited to the static case. To investigate if the kernel has discontinuities at finite frequency the discussion above must be generalized to an ensemble which allows the number of particles to change in time. To this end, we consider the following statistical operator $\hat{\rho}(t)=\sum_k\alpha_k(t)|\Psi_{k}(t)\ket\bra\Psi_{k}(t)|$,
where $\a_k(t)$ are given time-dependent coefficients whose sum is equal to 1 and $|\Psi_k(t)\ket$ is the many-body state of $k$ particles at time $t$. It is possible to prove a Runge-Gross-like theorem for this ensemble which allows us to define the XC potential as a functional of the ensemble density. In this way, the functional derivative $\d v_\xc(\vr t)/\d n(\vr' t')$ contains no arbitrariness, but leaves the possibility for a discontinuity of the form $f^+_\xc(\vr,\vr';t-t')=f^-_\xc(\vr ,\vr';t-t')+g_\xc(\vr ;t-t')+g_\xc(\vr';t-t')$. If we vary $v_\xc(\vr t)$ with respect to the time-dependent number of particles $N(t')$ and evaluate the derivative at the ground state with $N=N^{+}_0$ an expression for the frequency-dependent discontinuity $g_\xc(\vr,\w)$ can be derived
\bea
\left.\frac{\d v_\xc(\vr t)}{\d N(t')}\right |_{n^+_0}=\int d\vr' f^-_\xc(\vr,\vr', t-t')f^+(\vr')\nonumber\\
+\int d\vr' g_\xc(\vr', t-t')f^+(\vr')+g_\xc(\vr, t-t'),
\eea
where we again have used the chain rule. Below we will see with a numerical example how the frequency dependence modifies the discontinuity making the divergency stronger than in the static case and allows for a correct description of CT excitations in a combined donor-acceptor system. 
This will be illustrated in the time-dependent (TD) EXX approximation. For a derivation and analysis of the EXX kernel we refer the reader to Ref. \cite{hvB3,gorexx}.
From now on the subscript $\xc$ will be replaced by $\x$ to denote quantities in the TDEXX approximation.
\begin{figure}[t]
\includegraphics[width=8.5cm, clip=true]{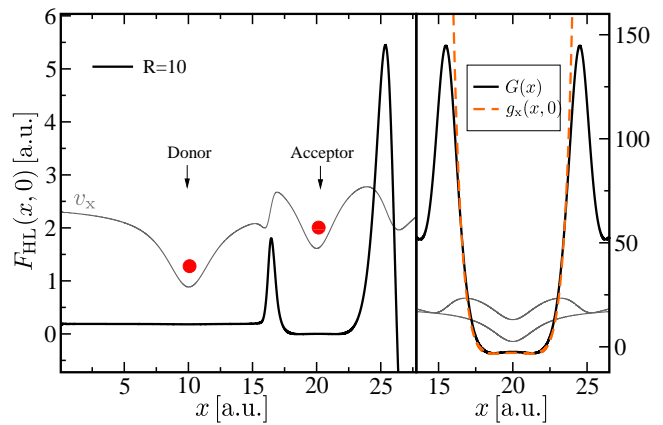}\\
\caption{Left: EXX potential and kernel for a heteronuclear system of four electrons. Right: the discontinuity $g_\x(x,0)$ calculated from Eq. (\ref{muxcdeltan}) for the isolated two-electron subsystem as well as $G(x)$ for $N=2.0001$. Note that the potentials have been rescaled and shifted for better visibility.}
\label{ediag}
\end{figure}
\section{Results} 
In the single-pole (SP) approximation \cite{pgg96} of TDDFT the XC correction to the KS excitation energy $\w_q$ is given by twice the matrix element $\bra q|f_{\xc}(\w)|q\ket=\int d\vr d\vr' \Phi_q(\vr) f_\xc(\vr,\vr',\w) \Phi_q(\vr')$ at $\w=\w_q$, where the index $q=ia$ corresponds to an arbitrary excitation. 
In Ref. \cite{gs99} it has been shown that in TDEXX $2\bra q|f_{\x}(\w_q)|q\ket=\bra a|\S_\x-v_\x|a\ket-\bra i|\S_\x-v_\x|i\ket-\bra aa|v|ii\ket$,
where $v$ is the Coulomb interaction. Considering a CT excitation between HOMO ($i={\rm H}$) and LUMO ($a={\rm L}$) we see that the last term goes as $1/R$. Setting, as usual \cite{gkkg}, $\bra {\rm H}|\S_\x-v_\x|{\rm H}\ket=0$ and using the previously mentioned result $\bra {\rm L}|\S_\x-v_\x|{\rm L}\ket=\D_\x$, we can deduce that \cite{ctneep}
\bea
\w_{\rm CT}&=&\w_{\rm HL}+2\bra {\rm HL}|v+f_{\x}(\w_{\rm HL})|{\rm HL}\ket\nn\\
&&\to\w_{\rm HL}+\D_{\x}-1/R.
\label{asymp}
\eea
The kernel $f_\x$ thus produces a finite correction if evaluated at $\w_{\rm HL}$ and yields exactly the results corresponding to first order G\"orling-Levy perturbation theory \cite{glper,gs99,spagk,spkf}. In the following we will see that it is the discontinuity of the kernel that yields this correct result. We have deliberately used the SPA and not the full solution of Casida equations in conjunction with the EXX kernel, a procedure which would imply the inclusion of higher orders in the explicit dependence on the Coulomb interaction. Our motives are to study an exact property of the kernel well captured in the SPA of TDEXX but may be subject to errors inherent to the approximation when including higher orders.  

We model \cite{jes88} a stretched diatomic molecule in terms of 1D atoms described by $Q/\sqrt{(x-x_0)^2+1}$, where $x_0$ is the location of the atom and $Q$ is the nuclear charge, and replace everywhere the Coulomb interaction $v$ with a soft-Coulomb interaction $1/\sqrt{(x-x')^2+1}$. We study two different systems, one ionic and one neutral system. In the ionic system the discontinuity is important already at the level of the XC potential whereas in the neutral system the discontinuity appears only in the XC kernel.
\subsection{Ionic system}
In the first example we study an ionic system and set $Q=2$ on the left atom (donor) and that of the right atom (acceptor) to $Q=4$ and solve the ground-state KS problem with 4 electrons. In the ground state at internuclear separation $R=10$ a.u. we find two electrons on each atom and in Fig. \ref{ediag} (left panel) the EXX potential is displayed (fade line) in arbitrary units.  Two steps are clearly visible, one between the atoms and another 
one on the right side of the acceptor. As a consequence, $v_\x$ is shifted upwards in the acceptor region placing the KS LUMO of the isolated acceptor above the HOMO of the isolated donor. This implies that the KS affinity of the acceptor becomes closer to the true affinity. In the same figure and panel we also display the quantity, 
$
F_{\rm HL}(x,\w)=\int dx' f_\x(x,x',\w)\Phi_{\rm HL}(x'),
\label{intker}
$
at $\w=0$. The function $F_{\rm HL}$ is seen to have peaks in correspondence with the steps of $v_\x$ and is shifted downwards over the acceptor with respect to the donor. Despite the fact that $\Phi_{\rm HL}$ tends to zero as $R$ increases the peaks of $F_{\rm HL}$ become sharper and higher, and the shift increases in size. The right panel in the same figure shows the function $G(x)=\int  dx' f_\x(x,x',0)f(x')$, accessible from Eq. (\ref{muxcdeltan}), for the isolated acceptor when $N=2.0001$ (full line), as well as the discontinuity $g_\x(\vr,0)$ in the limit $N=2^+$ (dashed line). The potentials $v_\x$ are also shown (fade lines) calculated from the same ensembles, i.e., $N=2.0001$ and $N=2$. The function $G(x)$ has peaks whose positions follow the steps of $v_\x$ and whose hight increases as $N$ approaches $2^+$. In the same limit, also the difference between the $G(\infty)$ value and the value of $G$ in the central plateau-like region increases consistently with the fact that $\int dx \, G(x)f(x)$ has to remain finite (see Eq. (\ref{muxcdeltandeltan})). Eventually $G(x)$ turns into the discontinuity $g_\x(x,0)$ which diverges as $x\to \infty$, in agreement with our previous analysis. Notice that CEDA has not been made here. If we compare $F_{\rm HL}$ and $G$ from the different panels we see a very similar structure. We therefore conclude that the peaked structure of the kernel in the donor-acceptor system is just the discontinuity of the ensemble EXX kernel. As discussed above, most part of the CT excitation energy is contained already in the KS eigenvalue differences due to the step in $v_\x$. A more critical example is therefore the neutral system, studied below. 
\subsection{Neutral system}
\begin{figure}[t]
\includegraphics[width=7.5cm, clip=true]{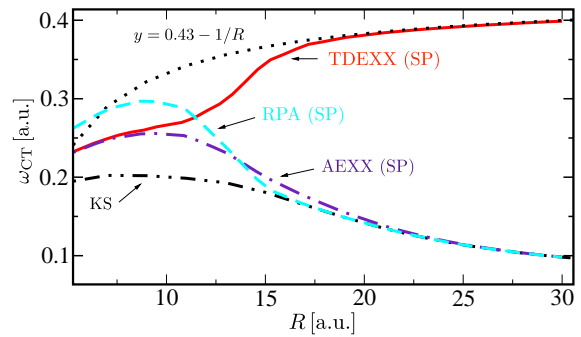}\\
\caption{CT excitation energies as a function of separation $R$ in different approximations. }
\label{excit}
\end{figure}

An example where the kernel needs to account for $\D_\x$ is the same system 
\begin{figure}[t]
\includegraphics[width=8.5cm, clip=true]{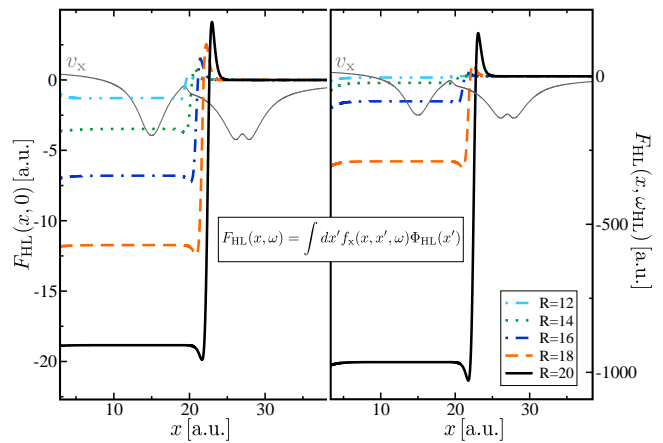}\\
\caption{Same system as in Fig. \ref{ediag} but with six electrons. Left: AEXX kernel. Right: TDEXX kernel at $\w_{\rm HL}$. Note that the potentials have been rescaled for better visibility.}
\label{ediag2}
\end{figure}
but with 6 electrons, i.e., a neutral system. The ground-state has 2 electrons on the left atom (acceptor) and 4 electrons on the right atom (donor). In Fig. \ref{ediag2} we plot the EXX potential (fade line) for $R=12$ a.u. and a step-like structure between the atoms can be observed. However, as $R$ is increased the step reduces in size and eventually goes to zero. In the left panel we plot $F_{\rm HL}$ at $\w=0$, i.e., in the adiabatic (AEXX) approximation, and for different separations $R$. Again, we find a peak structure in the kernel between the donor and the acceptor as well as a shift that increases exponentially with $R$. In this case, compared to the previous, the peaks are less pronounced but the step due to the plateau is much larger. Evaluating the kernel at the first KS CT excitation frequency increases the exponential growth of the step by around a factor of two (right panel). Thus, whereas the overall shape remains unaltered the magnitude of the shift is strongly influenced by the frequency dependence. This fact plays a crucial role in the description of the CT excitation for this system. We notice here that even if the step in $v_\xc$ disappears in the dissociation limit the step in $f_\xc$ remains. This is not a contradiction as the discontinuity might show up only in the second derivative. Fig. \ref{excit} illustrates the behavior of the SP CT excitation energy as a function of $R$ for the system of Fig. \ref{ediag2} in four different approximations. In TDEXX with the correction $\bra {\rm HL}|f_{\x}(\w_{\rm HL})|{\rm HL}\ket$ we find that the divergency of the kernel over the acceptor exactly compensates for the decreasing overlap $\Phi_{\rm HL}$, thus yielding a finite value as $R\to \infty$ as well as the right $1/R$ asymptotic behavior as it should according to Eq. (\ref{asymp}). In the adiabatic case we find instead that $\bra {\rm HL}|f_{\x}(0)|{\rm HL}\ket$ tends to 0 as $R\to \infty$, although reproducing the fully frequency dependent result up to around $R=8$.
We notice that even if the kernel is very large over the acceptor it will not affect the excitations which are localized there since any constant will vanish by the fact that $\Phi_q$ integrates to zero. Thus only excitations involving a transfer of charge from one atom to the other will be influenced by the discontinuity. 
\section{Conclusions}
In conclusion we have analyzed the discontinuity of the XC kernel of an ensemble with time-dependent particle numbers. In a combined system of two atoms we have seen that the divergency of the discontinuity as $r\to\infty$ can generate a kernel which diverges in the dissociation limit, and thus compensate for the vanishing overlap of acceptor and donor orbitals. This feature is crucial for the description of CT excitations but may also be important whenever there are excitations for which the KS orbital overlap is too small to give a correction. 


\end{document}